\PassOptionsToPackage{table,xcdraw}{xcolor}
\documentclass[sigconf,natbib=true,anonymous=false]{acmart}
\graphicspath{{./images/}}

\usepackage{pifont}% http://ctan.org/pkg/pifont

\usepackage{multirow}
\usepackage{multicol}
\usepackage{subfigure}
\usepackage{bm}
\usepackage{color}
\usepackage{adjustbox}

\usepackage{enumitem}
\usepackage{bbm}
\usepackage{graphicx}
\usepackage{mathtools}
\usepackage{bbm}
\usepackage{balance}
\usepackage{graphics}
\usepackage{bbding}
\usepackage{algorithm}
\usepackage{algpseudocode}
\usepackage{booktabs}
\usepackage[normalem]{ulem}
\useunder{\uline}{\ul}{}
\usepackage[table,xcdraw]{xcolor}

\AtBeginDocument{%
  \providecommand\BibTeX{{%
    \normalfont B\kern-0.5em{\scshape i\kern-0.25em b}\kern-0.8em\TeX}}}

\copyrightyear{2024} 
\acmYear{2024} 
\setcopyright{acmlicensed}
\acmConference[WWW '24]{Proceedings of the ACM Web Conference 2024}{May 13--17, 2024}{Singapore, Singapore} \acmBooktitle{Proceedings of the ACM Web Conference 2024 (WWW '24), May 13--17, 2024, Singapore, Singapore} \acmDOI{10.1145/3589334.3645661} 
\acmISBN{979-8-4007-0171-9/24/05}
\begin{document}

\title{Is Contrastive Learning Necessary? A Study of Data Augmentation vs Contrastive Learning in Sequential Recommendation}
% \title{The Data Augmentation Dilemma: Do We Really Need Contrastive Learning for Sequential Recommendation?}
% \title{The Data Augmentation Dilemma: Is Contrastive Learning Truly More Effective than Simple Data Augmentation in Sequential Recommendation?}
\author{Peilin Zhou}
\authornote{Both authors contributed equally.}
\affiliation{%
  \institution{Hong Kong University of Science and Technology (Guangzhou)}
  \country{}
  }
  \email{zhoupalin@gmail.com}

\author{You-Liang Huang}
\authornotemark[1]
\affiliation{%
  \institution{Hong Kong University of Science and Technology (Guangzhou)}
  \country{}
  }
  \email{yhuang142@connect.hkust-gz.edu.cn}

\author{Yueqi Xie}
\affiliation{%
  \institution{Hong Kong University of Science and Technology}
  \country{}
  }
\email{yxieay@connect.ust.hk}

\author{Jingqi Gao}
\affiliation{%
  \institution{Upstage}
  \country{}
}
\email{mrgao.ary@gmail.com}

\author{Shoujin Wang}
\affiliation{%
  \institution{University of Technology Sydney}
  \country{}
  }
  \email{shoujin.wang@uts.edu.au}

\author{Jae Boum Kim}
\affiliation{%
  \institution{Hong Kong University of Science and Technology}
  \country{}
  }
  \email{jbkim@cse.ust.hk}

\author{Sunghun Kim}
\authornote{Corresponding author.}
\affiliation{%
  \institution{Hong Kong University of Science and Technology (Guangzhou)}
  \country{}
  }
  \email{hunkim@cse.ust.hk}
\renewcommand{\shortauthors}{Peilin Zhou et al.}

\begin{abstract}
Sequential recommender systems (SRS) are designed to predict users' future behaviors based on their historical interaction data. Recent research has increasingly utilized contrastive learning (CL) to leverage unsupervised signals to alleviate the data sparsity issue in SRS. In general, CL-based SRS first augments the raw sequential interaction data by using data augmentation strategies and employs a contrastive training scheme to enforce the representations of those sequences from the same raw interaction data to be similar. Despite the growing popularity of CL, data augmentation, as a basic component of CL, has not received sufficient attention. 
This raises the question: \textit{Is it possible to achieve superior recommendation results solely through data augmentation?} 
To answer this question, we benchmark eight widely used data augmentation strategies, as well as state-of-the-art CL-based SRS methods, on four real-world datasets under both warm- and cold-start settings.
Intriguingly, the conclusion drawn from our study is that, certain data augmentation strategies can achieve similar or even superior performance compared with some CL-based methods, demonstrating the potential to significantly alleviate the data sparsity issue with fewer computational overhead.
We hope that our study can further inspire more fundamental studies on the key functional components of complex CL techniques. 
Our processed datasets and codes are available at \url{https://github.com/AIM-SE/DA4Rec}.

\end{abstract}


\maketitle
\section{Introduction}
\begin{figure}[t]
  \centering
  \includegraphics[width=\columnwidth]{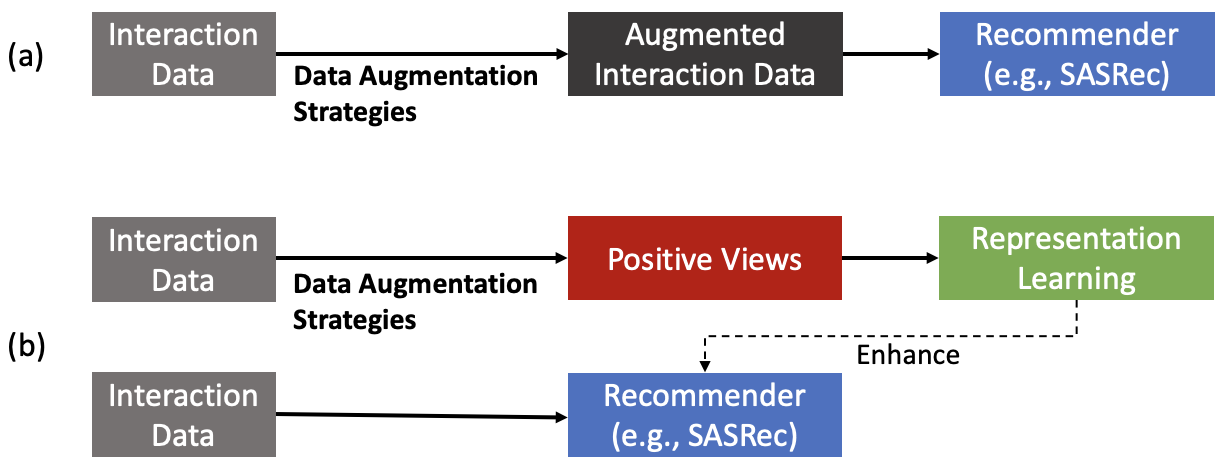}
  \caption{(a) Direct data augmentation for sequential recommendation; (b) Contrastive learning for sequential recommendation.}
\label{fig:intro}
\end{figure}
\begin{figure*}[t]
  \centering
\includegraphics[width=0.98\textwidth]{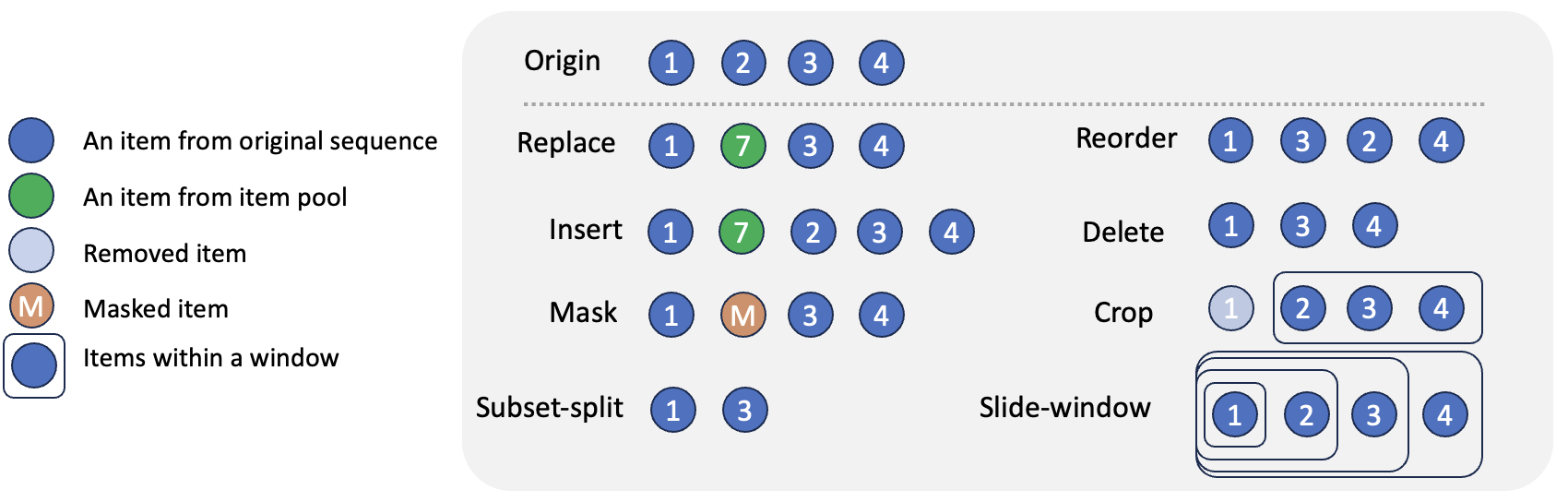}
  \caption{Eight widely used sequence-level data augmentation strategies.}
\label{fig:method}
\end{figure*}
Sequential recommender systems (SRS) play a crucial role in various domains, such as e-commerce~\cite{ecomm1, ecomm2, ecomm3,lei2023practical,zhang2023survey}, video~\cite{vid1, vid2}, music~\cite{mus1, mus2} and social media~\cite{soci1, soci2}. The goal of these SRS is to predict the next item that a user is likely to interact with based on his/her historical behavior~\cite{wang2022sequential,wang2019sequential}. One predominant obstacle in developing SRS is the data sparsity issue, where user-item interaction data is typically limited compared with a large number of users and items, leading to insufficient training signals to learn informative item representations for the downstream recommendations.

Recently, contrastive learning for recommendation has attracted increasing attention due to its remarkable capability to enhance item representations through the extraction of self-supervised signals from user-item interaction data. Consequently, various contrastive learning-based SRS, such as CL4SRec~\cite{cls4rec}, CoSeRec~\cite{coserec}, ICLRec~\cite{iclrec} and DuoRec~\cite{duorec}, have been proposed. The core idea of these methods can be summarized into two interrelated steps: (1) generating positive views and negative views through data augmentation strategies; and (2) minimizing (resp. maximizing) the distance between positive (resp. negative) views using a contrastive loss function (such as InfoNCE~\cite{infonce}). 
In these methods, data augmentation strategies are applied solely to the auxiliary tasks (shown in Fig.~\ref{fig:intro} (b)) designed for contrastive learning rather than directly applied to the recommendation task itself (shown in Fig.~\ref{fig:intro} (a)).
Such practice naturally raises two critical questions:\textit{ Can the performance of SRS be improved by solely relying on data augmentation (i.e., the first step) without using contrastive learning paradigm (i.e., the second step)? Whether existing CL-based SR methods consistently outperform direct sequence-level data augmentation?} 

Answering these questions is crucial for revisiting the role of data augmentation strategies in sequential recommendation tasks. However, so far, the direct application of data augmentation to mitigate the data sparsity issue in sequential recommendation has not received sufficient attention. 
Only one study~\cite{song2022DataAS} has explored the effects of four augmentation strategies on sequential recommendation. Nevertheless, it still suffers from following limitations:
First, the augmentation strategies compared are not comprehensive. For instance, strategies like reorder and delete, commonly used in contrastive learning for sequential recommendation (SR), can also be employed as standalone data augmentation methods.
Second, comparison with contrastive learning methods are not conducted, which is essential for revisiting the effectiveness of contrastive learning in the sequential recommendation research.
Third, no insight or analysis is provided on the factors that lead to varying performance results from different augmentation strategies.
Therefore, there is an urgent need to conduct a more systematic empirical study to thoroughly investigate the effectiveness of data augmentation in improving the performance of SRS.

To bridge this research gap, we conduct a comprehensive experimental study to compare the performance of SRS based on data augmentation only and that of SRS based on full contrastive learning. To be specific, we focus on investigating the effectiveness of eight popular sequence-level augmentation strategies: \textit{insert, replace, crop, delete, mask, reorder, subset-split, and slide-window}. Most of these augmentation strategies have been widely adopted in contrastive learning-based SRS over the past five years. 
Specifically, we decouple these sequence-level data augmentation strategies from contrastive learning methods and directly apply them to augment the training sequences. Both the original sequences and the augmented sequences are inputted together into backbone models (such as SASRec~\cite{sasrec}) for training. Afterward, we benchmark these eight sequence-level data augmentation strategies and three state-of-the-art contrastive learning-based SRS on four widely used datasets. We also simulate different cold-start scenarios to further evaluate the applicability of sequence-level data augmentation. Furthermore, we conduct in-depth analysis on the impact of the size of data augmentations, as well as the computational efficiency of various augmentation strategies and contrastive learning methods.

The experimental results demonstrate that, when using SASRec as the backbone, certain sequence-level augmentation strategies can achieve comparative or even superior performance compared to some contrastive learning-based SR methods, while requiring less training and inference time. 
This finding not only validates the feasibility of directly using sequence-level augmentation to alleviate data sparsity issue, but also suggests that the current research community might underestimate the effectiveness of simple sequence-level augmentation and overly emphasize the necessity of contrastive learning in sequential recommendation tasks.
Note that our claim does not negate the effectiveness of contrastive learning in recommendation tasks. In fact, on most datasets and metrics, specific contrastive learning based methods can
 still achieve the best performance. However, we observe that the benefits of these CL-based methods over direct data augmentation are not as pronounced as expected, particularly when considering computational efficiency. 
From this perspective, we believe that while contrastive learning is a popular means to alleviate the data sparsity issue, it may not be the only choice, i.e., it is not absolutely "necessary" and direct data augmentation strategies are also worth considering in practical scenarios.

The contribution of this paper can be summarized as follows:

\begin{itemize}
    \item We systematically benchmark eight popular sequence-level data augmentation strategies and three representative contrastive learning methods for sequential recommendation, providing insights into their performance and applicability.
    \item We explore the potential synergies in combining various data augmentation strategies to improve sequential recommendation performance.
    \item We find that employing specific sequence-level augmentation strategies can effectively mitigate the problem of data sparsity in sequential recommender systems. Moreover, these strategies typically demand less training time and consume smaller GPU memory compared to CL-based methods.
\end{itemize}
\label{intro}

\section{Problem Formulation}
Consider two sets, $\mathcal{U}$ for users and $\mathcal{I}$ for items.
For each user $u$ belonging to $\mathcal{U}$, their past interactions are recorded as a sequence $\mathcal{S}^{u} = [v_1^u, v_2^u, ..., v_{|\mathcal{S}^{u}|}^u]$. In this sequence, each $v^{u}_i$, which is an item from $\mathcal{I}$, represents the user's $i$-th interaction, ordered by time, and $|\mathcal{S}^{u}|$ is the total number of interactions for user $u$.
All users' interactions are collectively represented by $\mathcal{S} = \{\mathcal{S}^1, \mathcal{S}^2, \dots, \mathcal{S}^{|\mathcal{U}|}\}$, with $|\mathcal{U}|$ indicating the total number of users.
The purpose of sequential recommendation systems is to use the sequence of interactions $\mathcal{S}^u$ for a user $u$ to forecast the next item $v_{|\mathcal{S}^{u}|+1}^u$ from $\mathcal{I}$, with which this user is likely to engage at the next time step.

\section{Sequence-level Data Augmentation Strategies}
Our experiments include multiple common rule-based sequence-level data augmentation strategies. All of them can be viewed as operators that create augmented sequences by performing certain transformations to the original sequence.
Fig.~\ref{fig:method} illustrates how to augment each interaction sequence using these operators.
\subsection{Item Insert}
The "insert" action begins by selecting an insertion position, followed by the insertion of a chosen item from the item pool, resulting in an augmented sequence. 
For the user's interaction sequence \( \mathcal{S}^{u} \), let \( i \) be an insertion position, and \( t \) be a chosen item from the item set \( \mathcal{I} \). The augmented sequence \( \mathcal{S}^{u'} \) can be represented as:
\begin{equation}
\mathcal{S}^{u'} = [v_1^u, \ldots, v_{i-1}^u, t, v_i^u, \ldots, v_{|\mathcal{S}^{u}|}^u].
\end{equation}
 
\subsection{Item Delete}
The "delete" action randomly selects an item from the sequence for removal, thereby generating an augmented sequence.
For \( \mathcal{S}^{u} \) and a randomly chosen item at position \( k \), the augmented sequence \( \mathcal{S}^{u'} \) is:
\begin{equation}
\mathcal{S}^{u'} = [] v_1^u, \ldots, v_{k-1}^u, v_{k+1}^u, \ldots, v_{|\mathcal{S}^{u}|}^u].
\end{equation}

\subsection{Item Replace}
The "replace" action begins by selecting the item in the sequence that will be replaced, followed by selecting an item from the item pool to substitute the chosen item, resulting in an augmented sequence.
For \( \mathcal{S}^{u} \), let \( j \) be a position of the item to be replaced, and \( t' \) be a chosen item from \( \mathcal{I} \). The augmented sequence \( \mathcal{S}^{u'} \) is:
\begin{equation}
\mathcal{S}^{u'} = [v_1^u, \ldots, v_{j-1}^u, t', v_{j+1}^u, \ldots, v_{|\mathcal{S}^{u}|}^u].
\end{equation}
\subsection{Item Crop}
The "crop" action initially selects a cutoff position, from which a continuous series of items with a specified length are extracted as an augmented sequence.
Let \( c \) be the cutoff position and \( l \) be the length of the cropped sequence. The augmented sequence \( \mathcal{S}^{u'} \) is:
\begin{equation}
\mathcal{S}^{u'} = [v_c^u, v_{c+1}^u, \ldots, v_{c+l-1}^u].
\end{equation}
\subsection{Item Mask}
The "mask" action initially chooses an item from the sequence and subsequently masks the ID of the selected item using a predefined mask symbol.
Let \( m \) be the position of the chosen item from sequence \( \mathcal{S}^{u} \). If \( \mu \) is the predefined mask symbol, the sequence after masking can be given as:
\begin{equation}
\mathcal{S}^{u'} = [v_1^u, \ldots, v_{m-1}^u, \mu, v_{m+1}^u, \ldots, v_{|\mathcal{S}^{u}|}^u].
\end{equation}

\subsection{Item Reorder}
The "reorder" action initially selects a sub-sequence of a specific length and subsequently shuffles the order of the items within that sub-sequence. The sub-sequence and the remaining parts of the original sequence are then combined according to their original order, resulting in an augmented sequence.
Consider a sub-sequence of length \( r \) starting at position \( d \) from \( \mathcal{S}^{u} \). Let \( \text{shuffle}(x) \) denote a function that shuffles the order of the elements in \( x \). The augmented sequence \( \mathcal{S}^{u'} \) after shuffling this sub-sequence is:
\begin{equation}
\mathcal{S}^{u'} = [v_1^u, \ldots, v_{d-1}^u, \text{shuffle}(v_d^u, \ldots, v_{d+r-1}^u), v_{d+r}^u, \ldots, v_{|\mathcal{S}^{u}|}^u].
\end{equation}

\subsection{Subset Split}
Similar to the dropout mechanism~\cite{srivastava2014dropout}, for subset split method, each item \(v^{u}_i\) in the original sequence \(\mathcal{S}^u\) will be included in \(\mathcal{S}^{u'}\) with a probability of \(1-\theta\) and the probability of discarding is \(\theta\):
\begin{equation}
v^{u'}_i =
\begin{cases}
v^{u}_i, & P = 1-\theta \\
\text{discarded}, & P = \theta
\end{cases}
\end{equation}
Thus, the augmented sequence \(\mathcal{S}^{u'}\) is essentially a "subset" of the original sequence \(\mathcal{S}^u\), which can be mathematically represented as:
\begin{equation}
\mathcal{S}^{u'} = [v^{u'}_1, v^{u'}_2, \dots, v^{u'}_{|\mathcal{S}^{u'}|}].
\end{equation}
Note that the length of the augmented sequence \(|\mathcal{S}^{u'}|\) can vary due to the random discarding process and is likely to be less than or equal to \(|\mathcal{S}^{u}|\).

\subsection{Slide-window}
The slide-window strategy generates a new segment of the sequence and its subsequent item as the prediction target at each step by sliding a fixed-length window along the user's behavioral sequence \( \mathcal{S}^{u} \).  The window initiates sliding with its right edge positioned to the left of the first item, and the sliding process concludes when the right edge reaches the last item.
For a given window length \( L \), at each step \( t \), the cropped sequence of augmented items from \( \mathcal{S}^{u} \) is:
\begin{equation}
\mathcal{S}_t^{u'} = [v_t^u, v_{t+1}^u, \ldots, v_{t+L-1}^u],
\end{equation}
where the sliding process iterating until \( t+L-1 = |\mathcal{S}^{u}| \). 

Note that for each sequence \( \mathcal{S}_t^{u'} \) obtained through the slide-window, the training label also changes, typically to the next item \( v_{t+L}^u \) in the sequence. Consequently, the slide-window not only increases the quantity of training data but also adds complexity and diversity to the training process by creating new prediction targets for each sequence segment.
This distinctive characteristic has led to the widespread adoption of the slide-window as a data preprocessing step in both traditional SR models and CL-based SR models, setting it apart from other data augmentation strategies that diversify model training by altering or perturbing specific elements of the original sequence while preserving the same prediction target.

\begin{table}
    \small
	\caption{Statistics of the datasets after preprocessing.}
	\label{tab:datasets}
	\setlength{\tabcolsep}{0.6mm}{
	\begin{tabular}{lrrrr}
	\toprule
	Dataset  & Beauty & Sports & ML-1m & Yelp \\
	\midrule
	\# Users & 22,363 & 35,598 & 6,041 & 30,499\\
	\# Items & 12,101 & 18,357 & 3,407 & 20,068 \\
	\# Avg. Actions / User & 8.9 & 8.3 & 165.5 & 10.4\\
	\# Avg. Actions / Item & 16.4 & 16.1 & 292.6 & 15.8\\
	\# Actions & 198,502 & 296,337 & 999,611 & 317,182 \\
	Sparsity & 99.93\% & 99.95\% & 95.15\% & 99.95\%\\
	\bottomrule
	\end{tabular}
	}
\end{table}
\begin{table*}[t]
\centering
\caption{Comparison of data augmentation strategies and contrastive learning-based SR methods. All the results are reported as percentages for clarity and ease of reading. The best results are indicated in bold, while the runner-up results are marked with an underline.
}
\label{tab:overall_single}
\resizebox{\textwidth}{!}{%
\begin{tabular}{l|cccc|cccc|cccc|cccc}
\hline
 & \multicolumn{4}{c|}{Beauty} & \multicolumn{4}{c|}{Sports} & \multicolumn{4}{c|}{Yelp} & \multicolumn{4}{c}{ML-1m} \\ \cline{2-17} 
 & \multicolumn{2}{c}{Recall} & \multicolumn{2}{c|}{NDCG} & \multicolumn{2}{c}{Recall} & \multicolumn{2}{c|}{NDCG} & \multicolumn{2}{c}{Recall} & \multicolumn{2}{c|}{NDCG} & \multicolumn{2}{c}{Recall} & \multicolumn{2}{c}{NDCG} \\
\multirow{-3}{*}{Model} & @10 & @20 & @10 & @20 & @10 & @20 & @10 & @20 & @10 & @20 & @10 & @20 & @10 & @20 & @10 & @20 \\ \hline
SASRec$^{*}$ & 4.52±0.04 & 6.47±0.13 & 2.24±0.03 & 2.73±0.05 & 2.32±0.08 & 3.41±0.08 & 1.18±0.05 & 1.46±0.05 & 3.51±0.07 & 4.68±0.09 & 2.32±0.03 & 2.61±0.04 & 5.63±0.40 & 8.45±0.30 & 2.79±0.15 & 3.50±0.12 \\
+ subset-split & 4.78±0.08 & 6.98±0.14 & 2.40±0.02 & 2.95±0.03 & 2.61±0.08 & 3.89±0.07 & 1.32±0.07 & 1.65±0.07 & 3.77±0.04 & 5.17±0.11 & 2.43±0.07 & 2.78±0.05 & { 7.57±0.56} & { 11.09±0.60} & { 3.68±0.26} & { 4.57±0.26} \\
+ crop & \underline{5.28±0.10} & { 7.72±0.05} & { 2.61±0.05} & { 3.22±0.04} & { 2.86±0.03} & { 4.28±0.04} & { 1.41±0.03} & { 1.77±0.03} & 4.17±0.14 & { 5.79±0.13} & 2.69±0.05 & 3.10±0.03 & \underline{9.08±0.42} & \underline{13.51±0.48} & \underline{4.23±0.18} & \underline{5.35±0.19} \\
+ delete & 4.83±0.12 & 7.10±0.04 & 2.39±0.09 & 2.96±0.06 & 2.52±0.06 & 3.80±0.08 & 1.26±0.04 & 1.59±0.05 & 3.73±0.08 & 5.11±0.13 & 2.42±0.03 & 2.76±0.01 & 6.75±0.12 & 9.59±0.26 & 3.29±0.03 & 4.01±0.09 \\
+ mask & 4.26±0.06 & 6.29±0.21 & 2.07±0.05 & 2.58±0.08 & 2.13±0.04 & 3.17±0.04 & 1.06±0.03 & 1.32±0.03 & 3.28±0.11 & 4.38±0.13 & 2.30±0.05 & 2.58±0.04 & 6.05±0.10 & 8.86±0.08 & 2.87±0.03 & 3.58±0.04 \\
+ reorder & 4.67±0.08 & 6.87±0.08 & 2.28±0.06 & 2.83±0.06 & 2.45±0.05 & 3.67±0.08 & 1.22±0.01 & 1.53±0.02 & 3.60±0.07 & 4.95±0.10 & 2.42±0.07 & 2.76±0.05 & 5.84±0.19 & 8.54±0.33 & 2.8±0.08 & 3.47±0.10 \\
+ insert & 4.62±0.12 & 6.79±0.11 & 2.28±0.05 & 2.83±0.05 & 2.38±0.09 & 3.54±0.13 & 1.17±0.07 & 1.46±0.08 & 3.62±0.05 & 4.87±0.05 & 2.46±0.12 & 2.78±0.11 & 6.67±0.54 & 9.76±0.27 & 3.23±0.24 & 4.01±0.16 \\
+ replace & 4.26±0.09 & 6.20±0.09 & 2.08±0.05 & 2.57±0.04 & 2.05±0.07 & 3.07±0.09 & 1.01±0.03 & 1.26±0.04 & 3.23±0.06 & 4.22±0.05 & 2.33±0.04 & 2.57±0.04 & 5.90±0.15 & 8.64±0.16 & 2.79±0.05 & 3.48±0.04 \\
+ slide-window & \textbf{8.07±0.08} & \textbf{11.56±0.07} & \textbf{3.95±0.02} & \textbf{4.83±0.01} & \textbf{5.21±0.07} & \textbf{7.7±0.06} & \textbf{2.4±0.04} & \textbf{3.03±0.04} & \textbf{6.02±0.05} & \textbf{8.62±0.02} & \textbf{3.8±0.02} & \textbf{4.46±0.02} & \textbf{22.59±0.98} & \textbf{31.88±0.11} & \textbf{12.48±0.45} & \textbf{14.83±0.24} \\\hline
CL4SRec$^{*}$ & { 5.20±0.13} & \underline{7.87±0.15} & \underline{2.65±0.02} & \underline{3.32±0.03} & \underline{3.32±0.07} & \underline{5.10±0.09} & \underline{1.66±0.05} & \underline{2.11±0.05} & { \underline{4.70±0.08}} & \underline{6.42±0.13} & \underline{2.89±0.08} & \underline{3.33±0.06} & 5.65±0.23 & 8.51±0.27 & 2.73±0.11 & 3.45±0.16 \\
CoSeRec$^{*}$ & 4.72±0.06 & 7.01±0.07 & 2.31±0.04 & 2.89±0.04 & 2.71±0.06 & 4.12±0.03 & 1.33±0.03 & 1.68±0.01 & 4.03±0.07 & 5.53±0.10 & { 2.65±0.03} & { 3.03±0.03} & 5.83±0.21 & 8.47±0.29 & 2.82±0.09 & 3.49±0.11 \\
ICLRec$^{*}$ & 4.84±0.05 & 7.14±0.12 & 2.42±0.05 & 3.01±0.06 & 2.66±0.03 & 4.00±0.09 & 1.30±0.03 & 1.64±0.03 & 3.74±0.03 & 5.10±0.05 & 2.54±0.02 & 2.88±0.03 & 5.87±0.1 & 8.74±0.2 & 2.78±0.03 & 3.5±0.08 \\ \hline

\end{tabular}%
}
\end{table*}
\begin{table*}[t]
\centering
\caption{Comparison of slide-window (SW) combined with different data augmentation strategies or contrastive learning methods. All the results are reported as percentages for clarity and ease of reading. The best results are indicated in bold, while the runner-up results are marked with an underline.}
\label{tab:overall_multi}
\resizebox{\textwidth}{!}{%
\begin{tabular}{l|cccc|cccc|cccc|cccc}
\hline
 & \multicolumn{4}{c|}{Beauty} & \multicolumn{4}{c|}{Sports} & \multicolumn{4}{c|}{Yelp} & \multicolumn{4}{c}{ML-1m} \\ \cline{2-17} 
 & \multicolumn{2}{c}{Recall} & \multicolumn{2}{c|}{NDCG} & \multicolumn{2}{c}{Recall} & \multicolumn{2}{c|}{NDCG} & \multicolumn{2}{c}{Recall} & \multicolumn{2}{c|}{NDCG} & \multicolumn{2}{c}{Recall} & \multicolumn{2}{c}{NDCG} \\
\multirow{-3}{*}{Model} & @10 & @20 & @10 & @20 & @10 & @20 & @10 & @20 & @10 & @20 & @10 & @20 & @10 & @20 & @10 & @20 \\ \hline
SASRec$^{*}$ + SW & 8.07±0.08 & 11.56±0.07 & 3.95±0.02 & 4.83±0.01 & 5.21±0.07 & 7.7±0.06 & 2.4±0.04 & 3.03±0.04 & 6.02±0.05 & 8.62±0.02 & 3.8±0.02 & 4.46±0.02 & 22.59±0.98 & 31.88±0.11 & 12.48±0.45 & 14.83±0.24 \\
+ subset-split & \textbf{8.48±0.14} & 12.04±0.2 & \textbf{4.2±0.08} & \textbf{5.1±0.1} & {\ul 5.54±0.02} & {\ul 8.32±0.08} & \textbf{2.57±0.02} & \textbf{3.27±0.02} & 6.28±0.06 & 9.05±0.12 & 3.91±0.02 & 4.6±0.04 & 22.89±0.38 & 31.78±0.31 & 12.66±0.3 & 14.9±0.28 \\
+ crop & 7.13±0.02 & 10.24±0.06 & 3.59±0.02 & 4.38±0.02 & 4.4±0.01 & 6.68±0.04 & 2.07±0.01 & 2.64±0.01 & 5.23±0.07 & 7.38±0.08 & 3.44±0.04 & 3.98±0.03 & 20.59±0.21 & 29.55±0.22 & 11.35±0.12 & 13.61±0.1 \\
+ delete & {\ul 8.33±0.07} & 12.02±0.09 & {\ul 4.11±0.06} & 5.04±0.06 & 5.19±0.14 & 7.84±0.21 & 2.38±0.06 & 3.05±0.08 & 6.1±0.07 & 8.83±0.03 & 3.84±0.02 & 4.52±0.01 & {\ul 23.29±0.58} & {\ul 32.12±0.55} & {\ul 12.96±0.29} & {\ul 15.19±0.31} \\
+ mask & 7.5±0.16 & 10.9±0.08 & 3.66±0.07 & 4.52±0.05 & 4.5±0.02 & 6.82±0.03 & 2.05±0.02 & 2.64±0.03 & 5.74±0.13 & 8.29±0.15 & 3.64±0.06 & 4.28±0.07 & 22.93±0.25 & 32.05±0.17 & 12.75±0.09 & 15.06±0.1 \\
+ reorder & 7.95±0.07 & 11.53±0.07 & 3.88±0.05 & 4.78±0.05 & 4.96±0.06 & 7.53±0.2 & 2.26±0.05 & 2.91±0.08 & 6.03±0.04 & 8.64±0.06 & 3.79±0.02 & 4.45±0.03 & 23.15±0.02 & 32.11±0.2 & 12.9±0.17 & 15.16±0.2 \\
+ insert & 8.45±0.18 & {\ul 12.05±0.17} & 4.17±0.1 & {\ul 5.07±0.07} & 5.34±0.16 & 8.1±0.22 & 2.48±0.08 & 3.17±0.1 & 6.21±0.09 & 9.16±0.1 & 3.87±0.03 & 4.61±0.03 & 22.87±0.12 & 31.84±0.1 & 12.7±0.1 & 14.96±0.07 \\
+ replace & 7.25±0.09 & 10.46±0.11 & 3.52±0.02 & 4.33±0.02 & 4.3±0.13 & 6.53±0.05 & 1.96±0.05 & 2.52±0.02 & 5.33±0.07 & 7.76±0.2 & 3.44±0.05 & 4.05±0.09 & 22.58±0.21 & 31.6±0.54 & 12.62±0.3 & 14.9±0.39 \\ \hline
CL4SRec$^{*}$ + SW & 7.68±0.13 & 11.4±0.1 & 3.78±0.08 & 4.72±0.07 & 5.16±0 & 7.98±0.05 & 2.51±0.04 & {\ul 3.23±0.04} & 6.37±0.03 & 9.56±0.07 & 3.57±0.06 & 4.37±0.06 & 21.66±0.55 & 31.98±0.77 & 11.37±0.33 & 13.97±0.37 \\
CoSeRec$^{*}$ + SW & 8.24±0.04 & \textbf{12.07±0.18} & 3.99±0.08 & 4.96±0.03 & 5.5±0.05 & 8.26±0.03 & 2.52±0.02 & 3.21±0.01 & \textbf{6.93±0.11} & \textbf{10.08±0.1} & \textbf{4.13±0.06} & \textbf{4.91±0.06} & 21.85±0.41 & 31.48±0.88 & 10.82±0.7 & 13.25±0.57 \\
ICLRec$^{*}$ + SW & 8.28±0.05 & 12±0.05 & 3.93±0.02 & 4.87±0.02 & \textbf{5.65±0.01} & \textbf{8.34±0.02} & {\ul 2.55±0.01} & {\ul 3.23±0.01} & {\ul 6.78±0.07} & {\ul 9.92±0.08} & {\ul 4.02±0.02} & {\ul 4.8±0.02} & \textbf{24.04±0.35} & \textbf{32.78±0.23} & \textbf{13.71±0.25} & \textbf{15.91±0.31} \\ \hline
\end{tabular}%
}
\end{table*}

\section{Experiment}
In this section, we carry out extensive experiments to answer the following main research questions:
\begin{itemize}
    \item \textbf{RQ1:} How do different sequence-level augmentation strategies compare against state-of-the-art contrastive learning based SR methods?
    \item \textbf{RQ2:} How do sequence-level augmentation and contrastive learning methods perform in cold-start scenarios? 
    \item \textbf{RQ3:} How do sequence-level augmentation and contrastive learning methods perform under varying levels of item popularity?
    \item \textbf{RQ4:} Does the size of augmentations affect the performance of sequence-level augmentation?
    \item \textbf{RQ5:} How is the efficiency of sequence-level data augmentation compared to contrastive learning methods?
\end{itemize}

\subsection{Experimental Settings}
\subsubsection{Dataset.}
Experiments are carried out on four well-known benchmark datasets with diverse distributions: \textbf{Beauty} and \textbf{Sports} are derived from Amazon review datasets\footnote{http://jmcauley.ucsd.edu/data/amazon/}~\cite{DBLP:conf/sigir/McAuleyTSH15}; \textbf{Yelp}\footnote{https://www.yelp.com/dataset} is a renowned business recommendation dataset; \textbf{ML-1m}\footnote{https://grouplens.org/datasets/movielens/1m/} is a popular movie rating dataset containing 1 million ratings from 6,000 users on 4,000 movies. We preprocess these datasets uniformly following the methodology in ~\cite{sasrec,zhou2020s3,yuan2021icai, caser,yuan2019simple,coserec} by eliminating items and users with less than 5 occurrences. Table~\ref{tab:datasets} presents the dataset characteristics after preprocessing.

\subsubsection{Evaluation Metrics.}
In our experiment, we select Recall@$K$ and NDCG@$K$ as the evaluation metrics for different methods, where $K$ can be either 10 or 20. 
These two metrics are widely adopted in existing sequential recommendation research~\cite{sun2019bert4rec, Xie2022DIF, coserec, iclrec}. 
Regarding dataset partitioning, we utilize a leave-one-out approach: the final two items within each user interaction sequence are allocated to the validation and test sets, respectively, while the remainder of the items is utilized for model training.
To ensure fair comparison, we follow the suggestion of ~\cite{krichene2020sampled, dallmann2021case} to calculate the ranking results  over the complete item set rather than a sampled subset.

\subsubsection{Baseline Models.}
\label{baselines}
The performance of eight data augmentation strategies is evaluated based on SASRec~\cite{sasrec}. The details of these data augmentation strategies have been described in Sec. 3. 
In addition, we select three representative SR methods as our baselines, all of which are based on contrastive learning:
(1) CL4SRec~\cite{cls4rec}: The first work to employ contrastive learning for sequential recommendation, it utilizes three sequence-level augmentation methods for generating positive pairs.
(2) CoSeRec~\cite{coserec}: Different from CL4SRec, this contrastive learning approach further considers the associations between items to enhance the quality of the positive pairs.
(3) ICLRec~\cite{iclrec}:  A contrastive learning method that adopts clustering technique to capture users' latent intent.

\subsubsection{Implementation Details.}
We use RecBole~\cite{recbole}, a unified platform widely adopted in academic research, to implement and evaluate all of these baselines.
In RecBole, \textit{slide-window} is enabled by default and the augmented data is used for training both SASRec and CL baselines. To facilitate a unified and fair evaluation of different data augmentation strategies, we turn off the \textit{slide-window} for these models and adds an "$*$" after the model name to distinguish them from their original setting\footnote{
For example, SASRec$^{*}$ only uses the last item of the training instance as a supervision signal, while SASRec$^{*}$+SW can be seen as an approximation of the original SASRec, wherein each item in the training sequence serves as a label to guide model training.}.
The models are trained with the Adam optimizer for 300 epochs, employing a batch size of 1024 and a learning rate of 0.001. For Beauty, Sports, and Yelp datasets, the maximum sequence length is set to 50, while for the ML-1m dataset, it is set to 200 due to its longer average sequence length.
For attention-based methods, we conduct a grid search on hyper-parameters to identify the optimal combination. The searching space is: number of self-attention layers $\in \left\{2, 3\right\}$, number of self-attention heads $\in\left\{2, 4\right\}$,dropout rate on the embedding matrix and attention matrix $\in \left\{0.1, 0.2, 0.3, 0.5\right\}$, hidden size $\in \left\{64, 128, 256\right\}$ and embedding $\in \left\{64, 128, 256\right\}$. 
Regarding data augmentation approaches, the \textit{slide-window} has a length of 50 for Beauty, Sports, and Yelp, and 200 for ML-1m to accommodate the varying average sequence lengths.
Other hyperparameters for data augmentation are as follows: for \textit{insert}, \textit{replace}, \textit{delete}, and \textit{mask}, a single item is inserted, replaced, deleted, or masked; the dropout factor $\theta$ of subset split is set to 0.25; the length of sub-sequence in \textit{crop} and \textit{reorder} is set to 2.
The selection of augmentation positions and augmented items in the aforementioned data augmentation strategies is obtained through random sampling according to a uniform distribution. For the slide-window strategy, the size of augmentations depends on the original sequence length and window length. For other augmentation strategies, their size of augmentations is regarded as a hyperparameter $n$\footnote{Note that \( n \) counts both the original sequence and its \( n-1 \) augmentations for each individual sequence.}
. We discuss the impact of the size of augmentations $n$ in Sec.~\ref{size_of_da}.
All baselines and our backbone are carefully tuned on the used datasets for best performance. 
To ensure the reliability of the results, each baseline in the benchmark is trained 5 times with different random seeds, and the mean value and standard deviation (Mean ± std) are reported in main tables (Tab.~\ref{tab:overall_single} and Tab.~\ref{tab:overall_multi}).

\subsection{Overall Performance (RQ1)}
\subsubsection{Performance of single data augmentation strategy}
In this section, we explore the impact of eight sequence-level augmentation strategies on recommendation performance and compare them with three classic contrastive learning-based SR models. Specifically, for each sequence-level augmentation strategy, we select SASRec$^{*}$ as the backbone. Each instance in the training set undergoes augmentation twice using the corresponding strategy. 
The results are presented in Tab.~\ref{tab:overall_single}, and we draw the following observations:

\noindent\textbf{ Most sequence-level data augmentations can improve the performance of SASRec.} Among the eight augmentation strategies, \textit{slide-window} yields the best results, followed by \textit{crop}. Specifically, \textit{slide-window} achieves average relative performance improvements of \textbf{96.2\%} and \textbf{85.1\%} in terms of Recall@20 and NDCG@20, respectively, on datasets with shorter average sequence lengths (Beauty, Sports, and Yelp). Furthermore, on the ML-1m dataset with longer sequences,\textit{slide-window} demonstrates more significant improvements, with Recall@20 and NDCG@20 increasing by \textbf{2.8x} and \textbf{3.2x}. Conversely, \textit{mask} and \textit{replace} perform poorly as data augmentation methods, reducing the performance of SASRec on Beauty, Sports, and Yelp datasets. This can be attributed to the detrimental impact of noise introduced by these methods on model training, particularly in shorter sequences.

\noindent \textbf{Some sequence-level data augmentation strategies outperform contrastive learning-based SR models.} Among them, \textit{slide-window} performs better than all contrastive learning methods, while \textit{cropping} achieves performance close to or even surpassing that of some contrastive learning methods in most cases. 
It is worth noting that, on the ML-1m dataset with longer sequence lengths, most of the sequence-level data augmentation strategies can achieve performance comparable to or even superior to contrastive learning-based methods.

\subsubsection{Performance of combined data augmentation strategy}
Furthermore, we evaluate the performance of combining \textit{slide-window} with other sequence-level augmentation strategies or contrastive learning methods, and summarize the results in Tab.~\ref{tab:overall_multi}.
We observe that, in most cases, the performance of \textit{slide-window + crop/mask} /\textit{replace} is inferior to that of using \textit{slide-window} alone. However, the combination of \textit{slide-window} with the other four augmentation strategies, namely \textit{subset-split, delete, reorder, and insert}, leads to an improvement in recommendation performance, highlighting the synergistic effect between these strategies. Particularly, on the Beauty and Sports datasets, the \textit{slide-window + subset-split} achieves the highest performance among all augmentation combinations, with an average relative improvement of \textbf{6.0\%} in Recall@20 and \textbf{6.8\%} in NDCG@20 compared to using \textit{slide-window} alone.

Contrastive learning methods also exhibit notable performance improvements when integrated with the \textit{slide-window} strategy. For instance, on the Yelp dataset, the combination of CoSeRec and the \textit{slide-window} strategy outperforms the use of CoSeRec alone, achieving increases of \textbf{82.3\% }in Recall@20 and \textbf{62.1\%} in NDCG@20. Similarly, on the ML-1m dataset, the combination of ICLRec and the \textit{slide-window} strategy achieves the best performance in most cases, leading to improvements of \textbf{2.8x} in Recall@20 and \textbf{3.6x} in NDCG@20 compared to using ICLRec alone. These  results indicate that integrating the \textit{slide-window} strategy with various sequence-level augmentation techniques or contrastive learning-based SR methods can further boost recommendation performance. It is noteworthy that contrastive learning methods exhibit significant performance improvements over pure data augmentation methods only when employed in conjunction with the \textit{slide-window} strategy. Otherwise, contrastive learning methods demonstrate comparable or even inferior performance compared to certain data augmentation strategies. 
\begin{figure}[t]
  \centering
  \includegraphics[width=0.96\columnwidth]{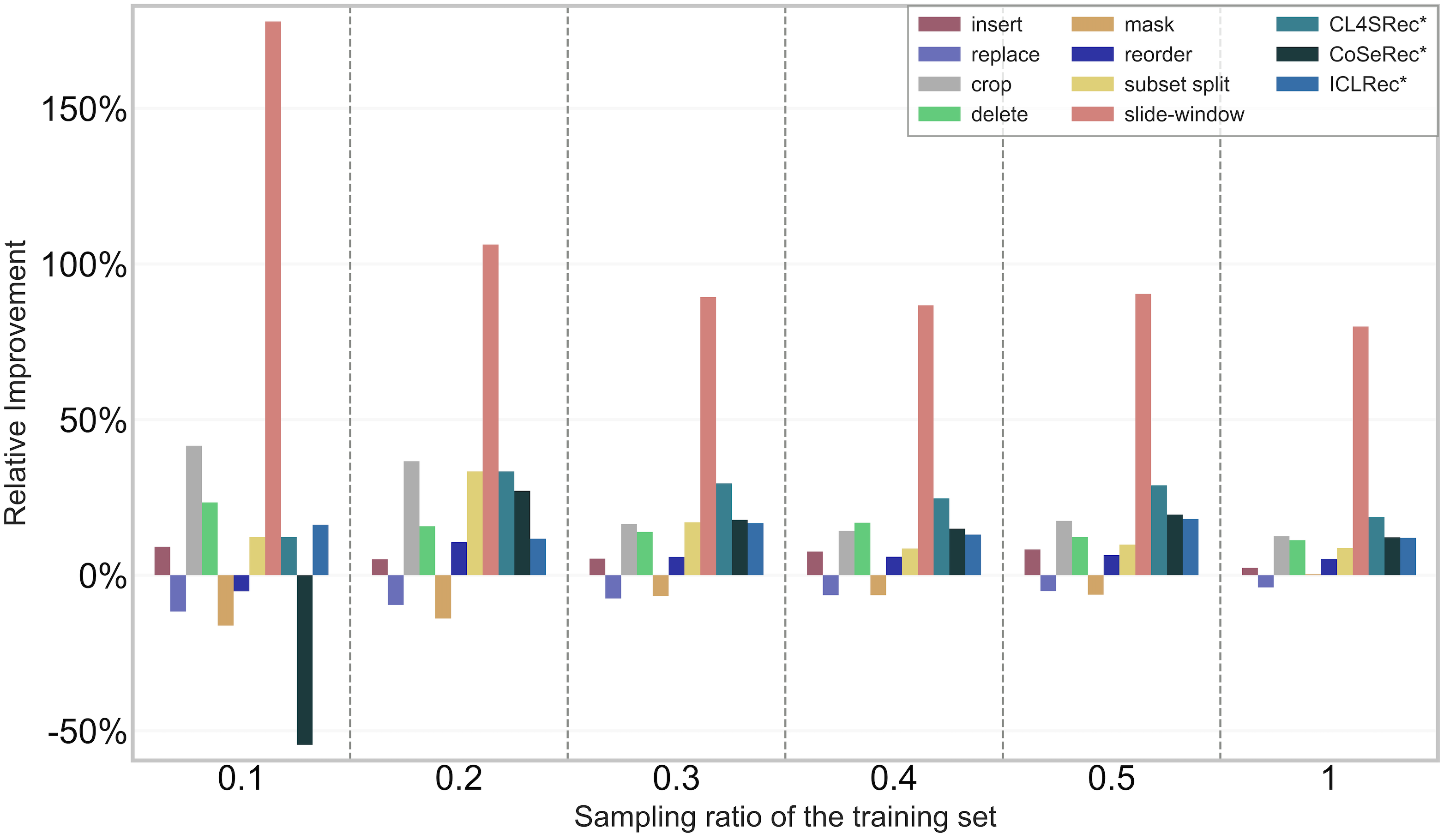}
  \caption{Performance improvements (Recall@20) of each data augmentation strategy over backbone model (i.e. SASRec) on Amazon Beauty dataset.}
\label{fig:relative_improvement}
\end{figure}
\begin{figure}[t]
  \centering
  \includegraphics[width=0.95\columnwidth]{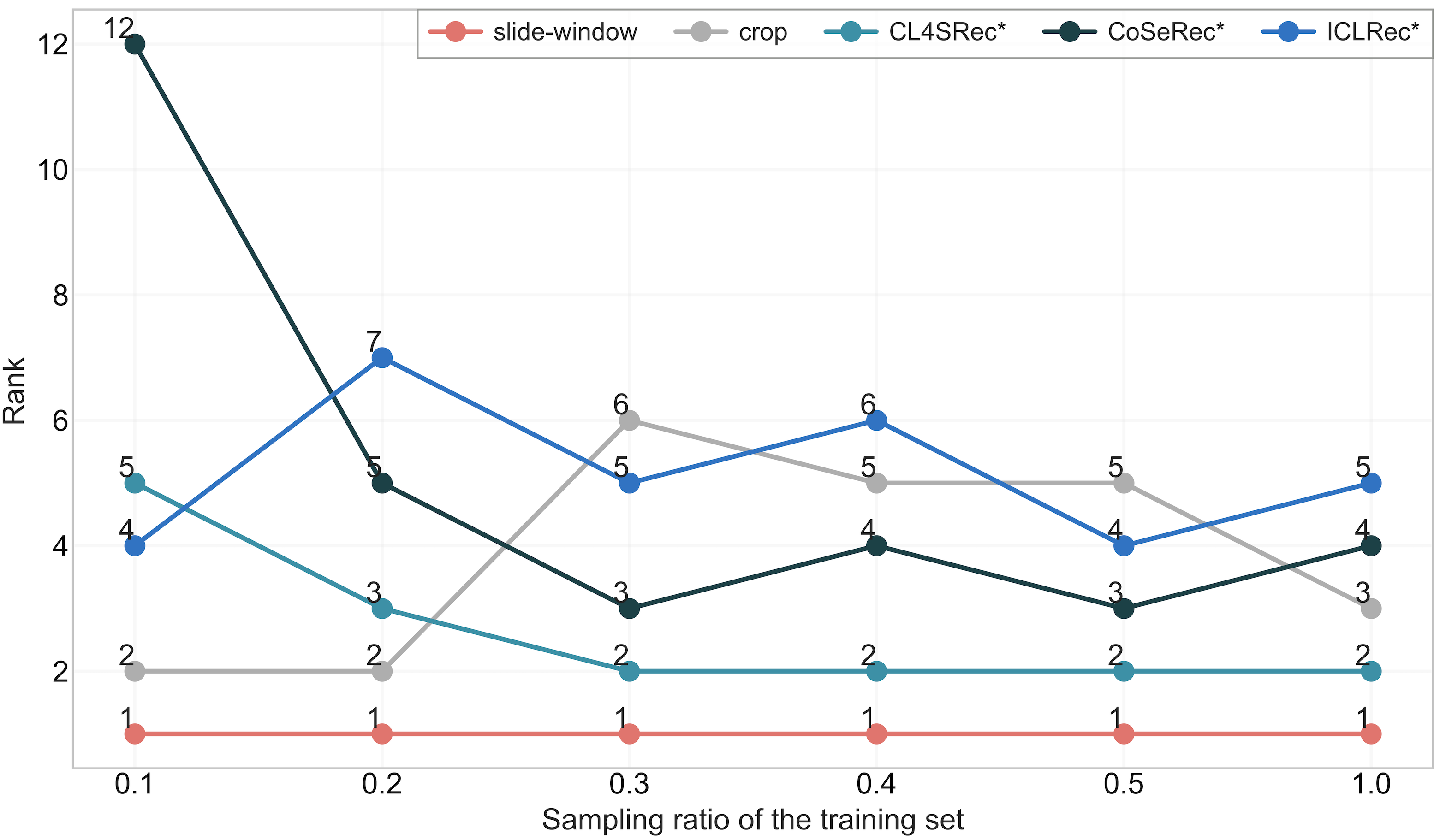}
  \caption{Performance ranking variations of two data augmentation strategies and three contrastive learning methods in various cold-start scenarios.}
\label{fig:rank}
\end{figure}
\begin{figure*}[t]
  \centering
  \includegraphics[width=0.90\textwidth]{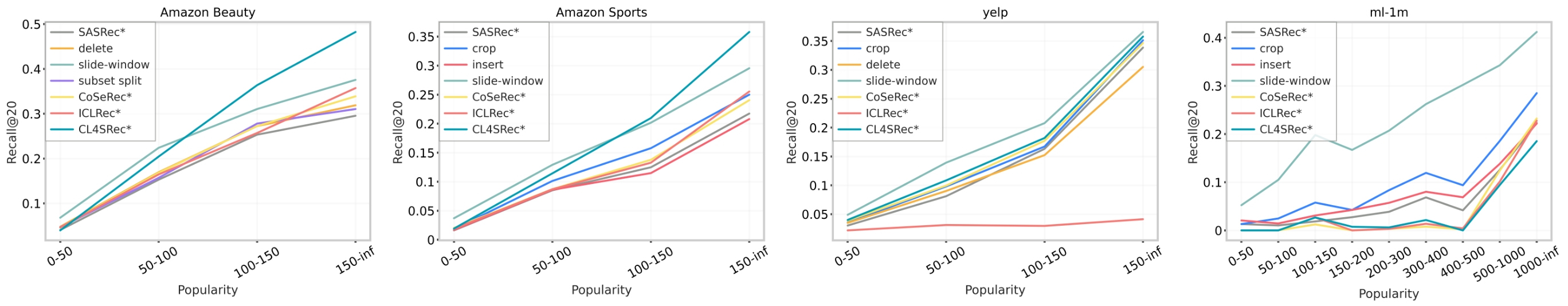}
  \caption{Performance comparison of different data augmentation and contrastive learning methods under different item popularity. For each dataset, we select the top-performing three data augmentation methods for comparison. Baseline denotes no augmentation is utilized.}
\label{fig:popularity}
\end{figure*}
\begin{figure*}[t]
  \centering
  \includegraphics[width=0.90\textwidth]{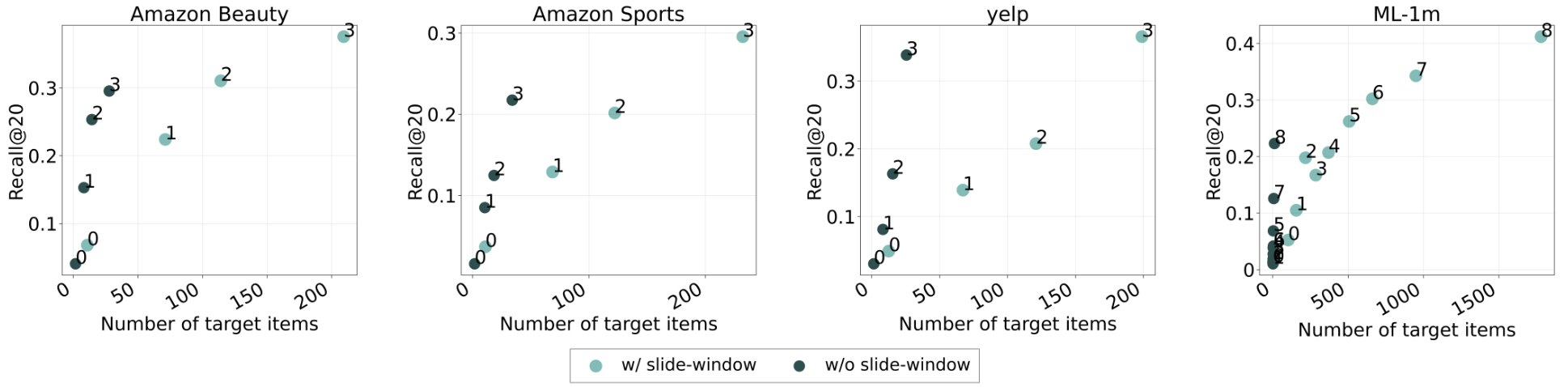}
  \caption{Performance comparison of SASRec with and without slide-window augmentation on target items with different popularity levels. The x-axis denotes the average number of target item from test set as the target item in the training set. The text next to the scatters denote which popularity interval it belongs to. For example, for the Amazon Beauty dataset, "0" denotes popularity interval 0-50, "1" represents interval 50-100, and so on.}
\label{fig:target_num}
\end{figure*}
\subsection{Cold-start Performance (RQ2)}
To compare different data augmentation strategies and contrastive learning methods in the cold-start scenario, we first simulate different levels of cold-start by randomly sampling the training data at proportions of [0.1, 0.2, 0.3, 0.4, 0.5]. Then, we apply each data augmentation strategy or contrastive learning method to the sampled training set and train the models using the augmented data. Finally, we assess the performance of all models on the original test set.  Fig.~\ref{fig:relative_improvement} illustrates the relative performance improvement ratios of the different methods compared to SASRec (no augmentation) on Amazon Beauty dataset.
We observe that in most cases, the less training data available, the more significant the relative performance improvement brought by different augmentation strategies.

Furthermore, we rank different contrastive learning methods and sequence-level augmentation methods (\textit{slide-window} and \textit{crop}) based on their performance (Recall@20) in various cold-start scenarios and depict the changes in their rankings in Fig~\ref{fig:rank}. It is evident that\textit{ slide-window} consistently outperforms other sequence-level augmentation methods and contrastive learning methods across all five cold-start settings. Interestingly, while \textit{crop} performs less favorably than contrastive learning methods on the original training set (i.e., sampling ratio equals 100\%), it surpasses all contrastive learning methods when the sampling ratios are reduced to 10\% and 20\%.
\subsection{In-depth Analysis (RQ3$\&$4$\&$5)}
\subsubsection{Performance under different levels of item popularity (RQ3)}
We compare the top-performing data augmentation methods and contrastive learning-based methods under varying item popularity on four benchmark datasets, using Recall@20 as the performance metric. 
In this work, the item popularity is defined as the frequency of each item as the target item in the training set. 
With such standard, the partition of sequences with different popularity is fixed and all data augmentation methods can be evaluated and fairly compared using the same test sequences. 
As shown in Fig~\ref{fig:popularity},  we can observe that all these methods, no matter direct data augmentation or contrastive learning-based ones, tend to perform better when the target item has higher popularity.
Actually, most items in recommendation datasets are less popular, and the higher the popularity, the fewer test data in that group. 
In low popularity groups (0-50 and 50-100), which contain most items, we observe significant performance improvement with the slide-window in all four datasets. For other direct data augmentation strategies, such as subset split and insert, they can also achieve comparable or even better performance on these low popularity groups. 

To further explore how simple data augmentation methods can have significant performance gains, we use the slide-window method as an example to test the effect of item popularity change on the model performance. 
In Fig~\ref{fig:target_num}, the ticks on the x-axis denote the average number of target item from test set as the target item in the training set. Addtionally, the text next to the scatters denotes which popularity interval it belongs to, and scatter with same number indicates they belong to the same target item popularity interval (e.g., scatters with text "0" in \textit{Amazon Sports} denotes they come from the first popularity interval, namely 0-50 of \textit{Amazon Sports} shown in Fig~\ref{fig:popularity}). As shown in Fig~\ref{fig:target_num}, the variation in model performance with the number of target items is approximately linear in the beginning, and then we can observe marginal utility of the performance improvement (in ML-1m dataset). This result, to some extent, indicates the model performance improvement brought by slide-window is obtaine by increasing the number of target item from test set as the target item in augmented training set.

\subsubsection{Impact of the size of data augmentations $n$ (RQ4)}
\label{size_of_da}
\begin{figure}[t]
  \centering\includegraphics[width=\columnwidth]{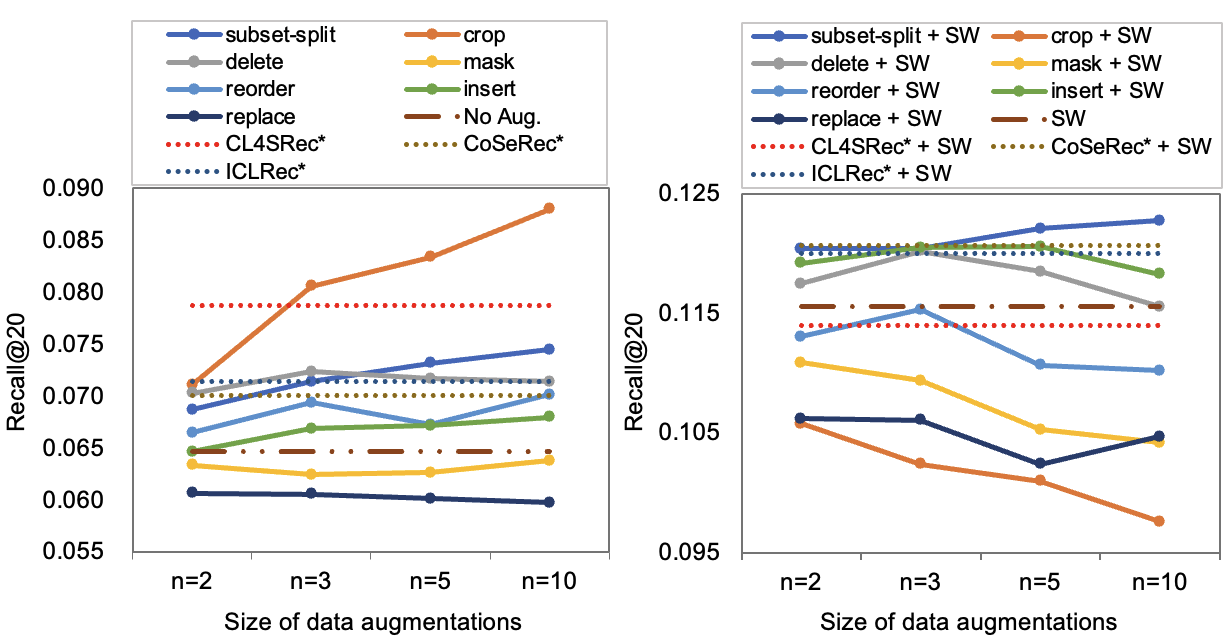}
  \caption{Impact of the size of data augmentations.}
\label{fig:augmentation_size}
\end{figure}

We compare the Recall@20 for different augmentation strategies when $n$ is set to 2, 3, 5, and 10. Fig ~\ref{fig:augmentation_size} presents the results on the Amazon Beauty dataset, where the left plot shows the results using a single sequence augmentation, and the right plot shows the results of slide-window augmentation combined with other augmentation strategies.
We observe that when using a single strategy, the Recall@20 for most augmentation strategies increases with the number of augmentations. However, when combined with slide-window augmentation, more than half of the augmentation strategies show a decrease in Recall@20 with larger size of augmentations. This could be attributed to the fact that excessive random augmentations introduce too much random noise during model training, resulting in the model's inability to accurately capture user interests.

\subsubsection{Comparison of efficiency (RQ5)}

As shown in Tab.~\ref{tab:time_complexity}, although the utilization of single or combined data augmentation strategies inevitably increases the volume of training data, their training time is still lower than contrastive learning methods. 
This is because contrastive learning methods usually introduce auxiliary tasks and complex positive view construction strategies, which increase computational overhead. Additionally, sequence-level augmentation strategies do not increase the model's inference time, as they only affect the training phase of the backbone model.

In terms of memory usage, our experiments, using Amazon Beauty as a case study with a batch size of 256, reveal substantial disparities among the methods. Direct data augmentation strategies, employing SASRec as the base model, maintains a memory usage of 0.36GB, consistent with SASRec's standalone implementation. In contrast, the three contrastive learning baselines require significantly higher memory allocations (CL4SRec: 0.78GB, CoSeRec: 1.37GB, and ICLRec: 2.35GB), highlighting the trade-off between improved performance and increased memory requirements in these approaches.

\begin{table}[t]
\centering
\caption{Comparison of training time and inference time.}
\label{tab:time_complexity}
\resizebox{1.0\columnwidth}{!}{%
\begin{tabular}{l|cc|cc|c}
\hline
 & \multicolumn{2}{c|}{Alone} & \multicolumn{2}{c|}{+SW} &  \\
\multirow{-2}{*}{Methods} & Recall@20 & Training time (s) & Recall@20 & Training time (s) & \multirow{-2}{*}{Test time (s)} \\ \hline
No aug. & 0.065 & {335.76} & 0.114 & {691.72} & 0.20 \\
slide-window & 0.114 & {691.72} & - & - & 0.20 \\
subset-split & 0.070 & {463.32} & 0.120 & {1232.58} & 0.20 \\
replace & 0.062 & {495.05} & 0.103 & {1443.56} & 0.20 \\
reorder & 0.069 & {453.19} & 0.114 & {1766.25} & 0.20 \\
mask & 0.063 & {510.61} & 0.109 & {1064.83} & 0.20 \\
insert & 0.068 & {431.12} & 0.118 & {2023.55} & 0.20 \\
delete & 0.071 & {467.31} & 0.119 & {2642.98} & 0.20 \\
crop & 0.077 & {539.55} & 0.105 & {1894.35} & 0.20 \\ \hline
CoSeRec$^{*}$ & 0.070 & {4650.78} & 0.117 & {13800.46} & {0.25} \\
CL4SRec$^{*}$ & 0.079 & {3321.12} & 0.114 & {8941.15} & 0.20 \\
ICLRec$^{*}$ & 0.071 & {1436.71} & 0.117 & {4055.04} & {0.26} \\ \hline
\end{tabular}%
}
\end{table}
\section{Related Work}  
\subsection{Sequential Recommendation}
Intensive studies about recommender systems of various real-world scenarios found that sequential behaviors are important signals to model user preferences. And many efforts have been devoted to leveraging sequential behaviors to better capture behavior patterns. The very early and the most intuitive method is adopting the Markov Chain assumption for sequential recommendation~\cite{rendle2010fpmc, he2016fusing}, where the next interaction is conditional on the past few interactions. Later, with the population of deep learning, many DL-based models were proposed to model sequential behaviors. GRU4Rec~\cite{gru4rec} is one of the most well-known SR models, of which Gated Recurrent Unit (GRU) is first introduced to model sequential behaviors. In addition, many other deep learning models were also introduced to seek better performance, such as Recurrent Neural Network (RNN)~\cite{li2017neural, wang2020intention}, Convolutional Neural Network (CNN)~\cite{tang2018personalized,jiang2023adamct}, Graph Neural Network (GNN)~\cite{chang2021seq,guo2022evolutionary,zhang2023can, wang2020modelling, wang2022exploiting}, and Multilayer Perceptron (MLP)~\cite{zhou2022filter}. Except for the aforementioned models, attention-based models have also been intensively studied and widely adopted in sequential recommendation tasks~\cite{kang2018self, sun2019bert4rec,zhang2023efficiently}. Besides, there are many interesting ongoing works focusing on other techniques like contrastive learning~\cite{cls4rec, coserec, iclrec,zhou2023equivariant,zhou2023attention}, reinforcement learning~\cite{xin2022self}, multi-interest learning~\cite{xie2023rethinking}, large language model~\cite{liu2023chatgpt, liu2023llmrec, zhou2023exploring} and relation awareness~\cite{ji2022seq}.
\subsection{Contrastive Learning for Recommendation}
Contrastive Learning (CL) aims to improve the quality of representations by reducing the distance between positive views generated from the same data instance while separating them from negative views in a latent space. In the field of sequential recommendation, sequence-level data augmentation or feature-level data augmentation is often used to create positive views, with augmented views of other data instances in the same training batch serving as negative views. For instance, CL4SRec~\cite{cls4rec} employed three sequence-level data augmentation techniques, namely cropping, masking, and reordering, to construct positive views. Subsequently, CoSeRec~\cite{coserec} proposed to generate robust augmented sequences based on item correlations. To mitigate the representation degradation, DuoRec~\cite{duorec} utilized feature-level augmentation based on dropout to better maintain semantic consistency between positive views. Despite these methods claiming that contrastive learning can significantly enhance the performance of recommender systems, they do not consider direct data augmentation as a baseline and thus cannot ascertain whether contrastive learning has a distinct advantage in mitigating data sparsity compared direct data augmentation.
\subsection{Data Augmentation for Recommendation}
Data augmentation is an effective method to improve the performance of DL-based models, particularly when the training data is scarce. In CV and NLP, data augmentation has drawn much attention and is widely adopted in model training. However, as for recommender systems, compared with CV or NLP, studies regarding data augmentation are still at a rather rudimentary stage.

For sequential recommendation tasks, basic data augmentation approaches create augmented sequences out of the original sequences themselves through simple transformations (\textit{e.g.,} crop, reorder), small perbulation (\textit{e.g.,} noise/redundancy injection, synonym replacement \cite{song2022DataAS}), or subset selection (\textit{e.g.,} slide-window \cite{tang2018personalized}, subset split \cite{tan2016rnn}, and item masking \cite{song2022DataAS}). Recent works regarding the aforementioned approaches focus on time-aware approaches, which better retain time coherence between the augmented sequences and the original ones, and further enhance the model's performance \cite{dang2023uniform, Petrov2022RSS}. Except for the aforementioned basic approaches, some data augmentation approaches choose to create highly plausible sequences by synthesizing and injecting/prepending fake samples into the original sequence \cite{huang2021MixGCF, liu2021ASReP, jiang2021sequential}, or modeling counterfactual data distribution \cite{zhang2021CauseRec, wang2021Counterfactual}. Apart from being applied in the sequential recommendation, data augmentation techniques are also applied in collaborative filtering to alleviate the data sparsity problem \cite{wang2019enhancing} or bypass negative sampling \cite{lee2021boot} during  training.

The work most related to ours is \cite{song2022DataAS}, which explored the impact of four augmentation strategies on sequential recommendation, namely noise injection, redundancy injection, item masking, and synonym replacement.
Different from it, our work benchmarked various direct data augmentation methods and contrastive learning methods, providing comprehensive analysis of the effectiveness of sequence-level data augmentation in sequential recommendation research, and offers insights into the improvements achieved.

\section{Conclusion and future work}
In this paper, we benchmark eight widely used sequence-level data augmentation strategies, as well as three state-of-the-art contrastive learning SR methods, on four real datasets under both full data and cold-start settings. The results reveal that the performance of SRS can be improved by solely relying on some data augmentation strategies without using contrastive learning paradigm. 
Moreover, certain sequence-level augmentation strategies can achieve comparative or even superior performance compared to some contrastive learning-based SR methods, while requiring less computational resources. 
In the future, we will extend the scope of benchmarking to include a broader range of data augmentation strategies and contrastive learning methods, providing theoretical justification for the effectiveness of direct data augmentation.

\newpage
\balance
\bibliographystyle{ACM-Reference-Format}
\balance
\bibliography{references}

\end{document}